\documentclass[aps,pra,showpacs,twocolumn]{revtex4}
\UseRawInputEncoding 
\usepackage{hyperref}
\usepackage{amsmath}
\usepackage{amssymb}
\usepackage{amsthm}
\usepackage{mathrsfs} 

\usepackage{graphicx}
\usepackage{xcolor}
\newcommand\mb{\boldsymbol}

\begin{document}

\title{Criterion for testing the covariance of physical laws \\ and Gordon optical metric}
\author{Changbiao Wang }
\email{changbiao\_wang@yahoo.com}
\affiliation{ShangGang Group, 10 Dover Road, North Haven, CT 06473, USA}

\begin{abstract}
The problem of covariance of physical quantities has not been solved fundamentally in the theory of relativity, which has caused a lot of confusion in the community; a typical example is the Gordon metric tensor, which was developed almost a century ago, and has been widely used to describe the equivalent gravitational effect of moving media on light propagation, predicting a novel physics of optical black hole.  In this paper, it is shown that under Lorentz transformation, a covariant tensor satisfies three rules: (1) the tensor keeps invariant in mathematical form in all inertial frames; (2) all elements of the tensor have the same physical definitions in all frames; (3) the tensor expression in one inertial frame does not include any physical quantities defined in other frames.  The three rules constitute a criterion for testing the covariance of physical laws, required by Einstein's principle of relativity.  Gordon metric does not satisfy Rule (3), and its covariance cannot be identified before a general refractive index is defined.  Finally, it is also shown as well that in the relativistic quantum mechanics, the Lorentz covariance of Dirac wave equation is not compatible with Einstein's mass--energy equivalence.
\end{abstract} 

\maketitle 

\section{Introduction}
\label{s1}
In the theory of relativity, physical quantities can be tensors, or their elements, or a combination of tensors and their elements.  A physical quantity is said to be covariant if the tensors associated with it are covariant under a change of reference frame. 
 
Laws of physics, such as Newton's laws and EM laws, are usually written in terms of physical quantities, and thus their covariance, required by Einstein's principle of relativity, is guaranteed by the covariance of tensors.

Unfortunately, the problem of covariance of tensors has never been solved fundamentally in the theory of relativity, which has caused a lot of confusion in the community; a typical example is the Gordon optical metric tensor \cite{r1}.  In this short paper, we take Maxwell equations and the four-momentum of a massive particle as examples to give the definition of covariance.  This definition consists of three rules, which constitute a criterion for testing the covariance of physical laws. 

\section{Properties of covariance of Maxwell equations}
\label{s2}  
In principle, the definitions of physical quantities used to construct a covariant theory of light propagation in moving media are supposed to be given in a general inertial frame, otherwise the covariance of the constructed theory is ambiguous. At least, one cannot identify  the covariance of the quantity without a definition given in a general frame.

As a first principle, Maxwell equations are the physical basis for descriptions of macro electromagnetic phenomena, and all other alternative approaches, including Lagrange formalism, must be confirmed by the Maxwell equations \cite[p.\,598]{r2}.  Thus any justified results that cannot be derived from Maxwell equations could constitute a challenge of the completeness of classical electromagnetic theory. 

Maxwell equations are also the perfect embodiment of Einstein's principle of relativity \cite{r3}, and they themselves contain the principle of the constancy of the speed of light (in free space) \cite{r4}.  According to Minkowski's electrodynamics of moving media, Maxwell equations are covariant under Lorentz transformation of two EM field-strength tensors \cite{r5}.  This covariance has three properties:
\begin{enumerate}
\item [(a)] Maxwell equations keep invariant in mathematical form in all Lorentz inertial frames.
\item[(b)] All physical quantities appearing in the equations have the same physical definitions in all frames.
\item[(c)] Maxwell equations in one frame do not include any physical quantities defined in other inertial frames.
\end{enumerate}
The above three properties do not contain each other. To ~put~ it~ simply, ~Maxwell equations ~are~ \emph{frame-independent} under Lorentz transformation of two EM field-strength tensors.  \\ \\ \\ \\ \\ \\ 

\section{Covariant form of four-vector of momentum-energy}
\label{s3} 
Another typical example for the covariance is the four-momentum of a massive particle, given by  
\begin{align}
&~\,m_0U^{\mu}=m_0\gamma(\mathbf{v},c)
\notag \\
&\textrm{(Non-covariant form)}
\notag \\
\notag \\
&~~~~~~~~~~=\frac{m}{\gamma}U^{\mu}
\notag \\
&\textrm{(Covariant form in relativity)}
\notag \\
\notag \\
&~~~~~~~~~~= \left(m\mathbf{v},\frac{mc^2}{c}\right)
\label{eq1} 
\\
&\textrm{(Covariant~form~in~3D space)}
\notag
\end{align} 
in a general $XYZ$ frame, where $m_0$  is the particle's mass in the \emph{particle-rest frame}\,---\,rest (proper) mass, $c$ is the vacuum light speed, $U^{\mu}=\gamma(\mathbf{v},c)$  is the four-velocity,  $m\,(=\gamma m_0)$ is the mass, $\gamma=(1-\mathbf{v}^2/c^2)^{-1/2}$, and $\mathbf{v}$  is the particle's velocity.  

The definition of $m_0U^{\mu}$ is in a non-covariant form. $m_0U^{\mu}$ has the same mathematical form and physical definition in all inertial frames, and complies with Properties (a) and (b), but it is not consistent with Property (c), because  $m_0U^{\mu}$  includes $m_0$ defined in the particle-rest frame, and thus  $m_0U^{\mu}$  cannot provide the frame-independent definitions of momentum and mass in a general frame. For example, $m_0U^{\mu}$  defines $m_0$ as the particle's mass in the particle-rest frame, but  $m_0U^{\mu}$  does not define the particle's mass $m$ in a general frame.

Properties (b) and (c) require that the definitions of particle's momentum and mass must be the same in all inertial frames and they do not include any quantities defined in other frames, namely
\begin{equation}
\textrm{Momentum}=\textrm{Mass}\times \textrm{Velocity}=m\mathbf{v}.
\label{eq2}
\end{equation}
Thus we have the mass transformation  $m=\gamma m_0$.

Obviously, the definition of four-momentum $(m/\gamma)U^{\mu}=(m\mathbf{v},mc^2/c)$ is frame-independent under its Lorentz transformation.  Especially, like Maxwell equations, it does not include any quantities of other frames; thus this definition complies with Property\,(c), in addition to Properties (a) and (b).

Note that $(m/\gamma)U^{\mu}$  is a covariant form in relativity, with $(m/\gamma)$ a \emph{covariant} scalar and   $U^{\mu}$ the four-velocity of the particle in a general $XYZ$ frame, while $(m\mathbf{v},mc^2/c)$  is a covariant form in 3D-space, with $m\mathbf{v}$  the momentum and $mc^2$  the energy of the particle in a general $XYZ$ frame; just like Maxwell equations have the two forms of covariance \cite[Footnote 7 there]{r6}.

\section{Is Gordon metric covariant?}
\label{s4} 
Unfortunately, Property\,(c) has been usually neglected in constructing a tensor in the community.  A typical example is Gordon optical metric tensor \cite{r1,r7,r8}, which was proposed in 1923 in a seminal paper by a German theoretical physicist Walter Gordon \cite{r1} to describe the equivalent gravitational effect of moving media on light propagation. Two decades ago, Leonhardt and Piwnicki generalized Gordon metric to investigate a non-uniformly moving medium \cite{r9}, and predicted a novel finding of optical black hole \cite{r10}, which has been arousing wide interest \cite{r11,r12,r13,r14,r15,r16}.  

The Gordon metric in a general $XYZ$ frame reads \cite{r1}: 
\begin{align}
\mathnormal{\Gamma}^{\mu\nu}=g^{\mu\nu}+(n_0^2-1)u^{\mu}u^{\nu},
\label{eq3}
\end{align} 
where $g^{\mu\nu}=\mathrm{diag}(-1,-1,-1,+1)$  is the Minkowski metric, $u^{\mu}=U^{\mu}/c=\gamma (\mb{\beta},1)$  with $\mb{\beta}=\mathbf{v}/c$  is the normalized four-velocity of the moving medium (namely the medium moves at $\mathbf{v}=\mb{\beta}c$  with respect to $XYZ$), and $n_0$  is the refractive index in the \emph{medium-rest frame}.

From Eq.\,(\ref{eq3}), we find that $\mathnormal{\Gamma}^{\mu\nu}$  is given in a general frame $XYZ$, but it includes $n_0$  which is a quantity defined in another frame (medium-rest frame).  Obviously, this is not consistent with Property\,(c).  In their works \cite{r9,r10}, Leonhardt and Piwnicki are supposed to provide the definition of the refractive index in a general frame in order to remove this inconsistency left by Gordon \cite{r1}, and make it expressed in a covariant form, just like Eq.\,(\ref{eq2}) leading to the \emph{frame-independent} four-momentum  $(m/\gamma)U^{\mu}=(m\mathbf{v},mc^2/c)$  shown above; unfortunately, the authors failed to do so.  Thus the covariance of  $\mathnormal{\Gamma}^{\mu\nu}$  cannot be identified, namely the covariance of Gordon optical metric $\mathnormal{\Gamma}^{\mu\nu}$ is ambiguous, depending on how to define the refractive index in a general inertial frame.

It ~should~ be indicated that Gordon ~defined ~the (proper) refractive index~ $n_0$~  by assuming ~that ~the medium is isotropic, observed in the medium-rest frame \cite{r1}.  However when a medium is moving, it becomes anisotropic in general \cite{r17}, and Gordon's way to define $n_0$  is not valid to define the refractive index in a general inertial frame. 

It also should be indicated that the way to define the refractive index in a general frame could essentially revise the descriptions of light propagation.  For example, in Wang's theory where the refractive index in a uniform medium is defined in a general inertial frame, propagation of light complies with Fermat's principle in all inertial frames \cite{r18}, while in Leonhardt's theory, it does only ``in the special case of a medium at rest'' \cite{r19}. 

\section{Conclusions}
\label{s5}
Based on above analysis, we now can make a general definition for Lorentz covariance of a tensor, as follows.  

Under Lorentz transformation, a tensor is said to be (Lorentz) covariant, if 
\begin{enumerate}
\item[(1)] the tensor keeps invariant in mathematical form in all inertial frames; 
\item[(2)] all elements of the tensor have the same physical definitions in all frames; 
\item[(3)] the tensor expression in one inertial frame does not include any physical quantities defined in other frames.
\end{enumerate}
The above three rules constitute a criterion for testing the covariance of physical laws.

As we know, Lorentz invariant (scalar) is the zeroth-rank tensor.  $(m/\gamma)$  given in Eq.\,(\ref{eq1}) fulfills all above three rules and it is a \emph{covariant} scalar, while the particle's rest mass $m_0$  is not a \emph{covariant} scalar because it at least breaches Rule (3).   $m_0$  is a physical quantity defined in the particle's rest frame, and it is the value of the \emph{covariant} scalar $(m/\gamma)$  in the particle's rest frame.

It should be noted that the definitions of momentum and mass in a general inertial frame, given by Eq.\,(\ref{eq2}), complies with Newton's law (momentum = mass $\times$ velocity), Einstein's mass--energy equivalence equation (energy = mass $\times ~c^2$) \cite{r4}, and principle of relativity (there must be the same definitions of mass and momentum in all inertial frames because inertial frames are equivalent and laws of physics are the same, no matter whether the space is filled \emph{partially or fully} with dielectric or materials). The mass $m$ in Eq.\,(\ref{eq2}) is also called ``relativistic (inertial) mass'' \cite[p.\,110]{r20} \cite[p.\,69]{r21} \cite[p.\,33]{r22}, and it is the total mass, a sum of the rest mass $m_0$  and the kinetic mass $(\gamma-1)m_0$.  The mass and energy are equivalent, and ``the mass of a body is a measure of its energy content''; this general conclusion was first drawn by Einstein in 1905 based on analysis of a special radiation process, where the body converts part of its \emph{rest} energy into the \emph{kinetic} energy of photons \cite{r4}.

Gordon optical metric does not fulfill Rule (3) either, and thus its covariance is ambiguous before a general refractive index is defined.  If no general definition of refractive index could be provided, then the equivalence of all inertial frames would be challenged, which is clearly in contradiction with Einstein's principle of relativity. Unfortunately, this subtle but apparently important basic problem in the Gordon's theory \cite{r1} has never been realized in the community. 

Traditionally, it has been argued that ``when a law of physics is written in terms of tensors, it will automatically be covariant under a change of reference frame" \cite{r23}.  However it should be emphasized that a tensor is not necessarily covariant.  The term ``tensor'' refers to the mathematical attribute of a physical quantity, while the ``covariance'' refers to the physical requirement from Einstein's principle of relativity, imposed on the physical quantity.  Gordon metric satisfies Lorentz transformation and it is a tensor, but its covariance is ambiguous, as shown in the present paper.  From this one can see that the traditional argument mentioned above is incorrect, because it is clearly disproved by the Gordon metric, in addition to the hugely controversial Abraham tensor \cite{r5}.  Note that (i) if Abraham tensor follows Lorentz transformation, then it does not fulfill Rules (1) and (2), or conversely, (ii) if Abraham tensor fulfills Rules (1) and (2), then it does not follows Lorentz transformation; otherwise its contradiction against two EM field-strength tensors is stimulated. 

Finally, it is interesting to point out that the definition of relativistic inertial mass for a particle, given by Eq.\,(\ref{eq2}), is consistent with Einstein's mass--energy equivalence equation, but it makes the rest mass $m_0$ not a \emph{covariant} scalar.  From this it follows that in relativistic quantum mechanics, Dirac wave equation for an electron is not covariant because the electron's rest mass appearing in Dirac equation is not a covariant scalar although Dirac took it to be in his proof \cite[p.\,258]{r24}.  If the rest mass $m_0$ in the four-momentum $m_0U^{\mu}=m_0\gamma (\mathbf{v}, c)$  is defined as a \emph{covariant} scalar, as argued in some respected textbooks \cite[p.\,289]{r25} and research works \cite{r26}, then it contradicts with Einstein's mass--energy equivalence equation.  Thus the Lorentz covariance of Dirac wave equation is not compatible with Einstein's mass--energy equivalence. \\


\newpage

\begin{figure} 
\includegraphics[trim=1.0in 1.0in 1.0in 1.0in, clip=true,scale=1.0]{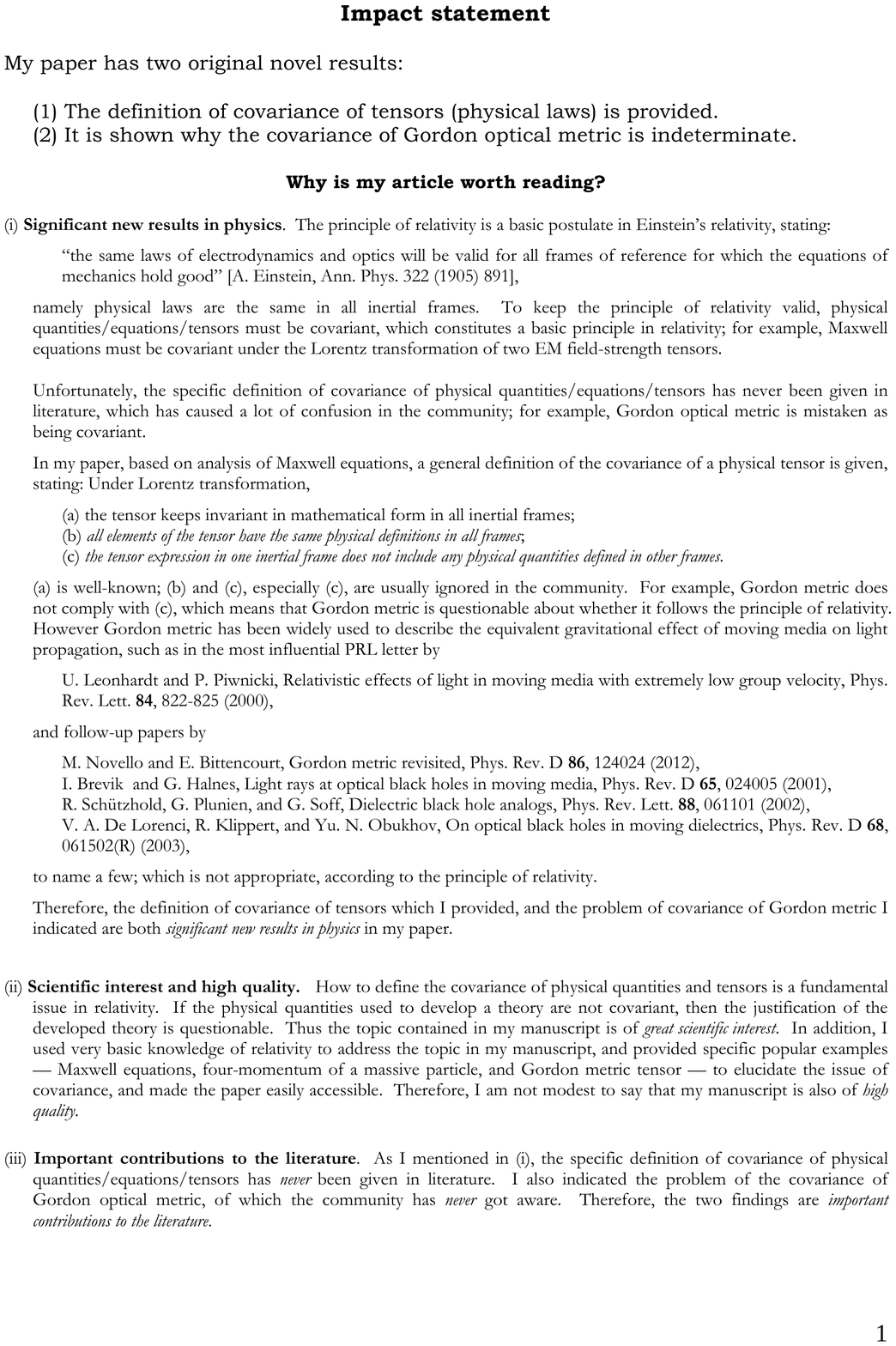}
\label{figM1}
\end{figure} 

\end{document}